\documentclass[aps,prb,twocolumn,showpacs,superscriptaddress]{revtex4}
\usepackage[cp866]{inputenc}
\usepackage{graphicx}
\usepackage{bm}

\newcommand{\beq}{\begin{equation}}
\newcommand{\eeq}{\end{equation}}

\newcommand{\mitr}{\mbox{\scriptsize MIT}}

\begin{document}

\title{Damping effects and the metal-insulator transition in
the two-dimensional electron gas}
\author{V.~A.~Khodel, M.~V.~Zverev,}
\affiliation{ Russian Research Centre Kurchatov Institute, Moscow,
123182, Russia}
\author{J.~W.~Clark}
\affiliation{cDonnell Center for the Space Sciences and
Department of Physics, Washington University, St.~Louis, MO 63130,
USA }

\date{\today}

\begin{abstract}
The damping of single-particle degrees of freedom in
strongly correlated two-dimensional Fermi systems is analyzed.
Suppression of the scattering amplitude due to the damping effects
is shown to play a key role in preserving the validity of the
Landau-Migdal quasiparticle picture in a region of a phase
transition, associated with the divergence of the quasiparticle
effective mass. The results of the analysis are applied to
elucidate the behavior of the conductivity $\sigma(T)$ of the
two-dimensional dilute electron gas in the density region where it
undergoes a metal-insulator transition.

\end{abstract}

\pacs{ 71.10.Hf,
71.27.+a
} \maketitle

A quantitative understanding of the damping of single-particle
excitations in a Fermi liquid (FL) is essential to the
determination of the resistivity, thermal conductivity, and other
kinetic properties of the system.  When the temperature dependence
of the properties of uncharged Fermi liquids is treated within
Landau theory, the decay rate $\gamma(\varepsilon)$ of
single-particle excitations at the relevant energies
$\varepsilon\sim T$ is given by \cite{pin}
\beq
\gamma(T)=W(M^*)^3T^2 \ .
\label{damp1}
\eeq
Here the effective
mass $M^*$  specifies the FL single-particle spectrum
$\xi(p)\equiv \epsilon(p)-\mu=p_F(p-p_F)/M^*$, where
$\epsilon(p)=\delta E_0/\delta n(p)$ and $\mu$ is the chemical
potential. The factor $W$ is proportional to the square of the
scattering amplitude $\Gamma$, suitably averaged over spins and
momenta of incoming and outgoing particles.

Reliable experimental data on the modification of FL properties
under variation of controllable variables (e.g., the density $n$)
exist only for two-dimensional (2D) Fermi systems, notably liquid
$^3$He and the electron gas.\cite{godfrin,saunders,krav,vitk,shash}
 Landau theory adequately
reproduces the behavior of these data in a broad density region,
{\it except} in the vicinity of the critical density $n_{\infty}$
where the effective mass diverges, and the spectrum $\xi(p)$
becomes flat.  This failure of FL theory is conventionally
attributed to a strong enhancement of the dimensionless damping
rate $r(T)=\gamma(T)/T$. Close to the critical point, $r(T)$
allegedly exceeds unity, invalidating the Landau-Migdal
quasiparticle picture.

Here we shall demonstrate that in actuality the parameter $r(T)$
remains rather small on both sides of the phase transition associated
with the divergence of the effective mass in the 2D system, and,
consequently, that the quasiparticle picture does apply.  We
then proceed to study kinetic phenomena within the quasiparticle
formalism, with particular attention to the metal-insulator transition
(MIT) occurring in the 2D electron gas in the density region where
the effective mass diverges.\cite{krav,vitk,shash}

Our analysis is based on the standard formula for the damping
rate,\cite{pin,trio}
$$
\gamma(\varepsilon) \sim-\sum_{{\bf p}_1,{\bf p}'}
\!\int\!\!\!\int \!\! W({\bf p},{\bf p}_1,{\bf p}',{\bf p}_1';
\varepsilon,\varepsilon_1,\omega)
F(\varepsilon,\varepsilon_1,\omega,T)
$$
\beq
{}\times
{\rm Im}\, G_R({\bf p}_1,{-}\varepsilon_1)
{\rm Im}\, G_R({\bf p}',\varepsilon-\omega)
{\rm Im}\, G_R({\bf p}_1',\omega{-}\varepsilon_1)
d\varepsilon_1 d\omega ,
\label{imsig}
\eeq
where ${\bf p},{\bf p}_1$ and ${\bf p}',{\bf p}_1'$ are
respectively the incoming and outgoing momentum pairs, and
$\omega{=}\varepsilon{-}\varepsilon'$. The function $W$ is given
by the sum of absolute squares of the scalar ($s$) and
spin-dependent ($a$) components $\Gamma_s$ and $\Gamma_a$ of the
scattering amplitude
$\Gamma=\Gamma_s+\Gamma_a{\bm\sigma}_1{\bm\sigma}_2$, while
$F(\varepsilon,\omega,\varepsilon_1,T)=\cosh({\varepsilon/2T})
[\cosh({\varepsilon_1/2T})\cosh((\varepsilon-\omega)/2T)
\cosh((\omega-\varepsilon_1)/2T)]^{-1}$ and $G_R$ is the retarded
Green function. In what follows we assume that the dependence of
the mass operator $\Sigma(p,\varepsilon)$ on $\varepsilon$ is not
crucial, and then
\beq
{\rm Im}\,G_R(p,\varepsilon)=
-\gamma(\varepsilon)/
[(\varepsilon-\xi(p))^2+\gamma^2(\varepsilon)] \ .
\label{imgr}
\eeq

To begin, we note that in the collision integral (\ref{imsig}),
all the quasiparticle energies must lie close to the Fermi
surface, so that $ |\xi(p)|\leq T$, $|\xi(p_1)|\leq T$, and
$|\xi(|{\bf p} -{\bf q}|)|\leq T$, $|\xi(|{\bf p}_1+{\bf q}_1|)|\leq T$,
 since, as we shall see, broadening of the
single-particle states is insignificant. In 2D, these conditions
are easily met, if (i) the momentum transfer $q=|{\bf p}-{\bf p}'|$
in the longitudinal particle-hole channel is small, i.e.,
$q\leq q_c(T)=T \left(dp/d\xi\right)_T\sim TM^*/p_F$, or
equivalently, if (ii) the momentum transfer $q_1=|{\bf p}-{\bf p}_1'|$
in the transverse particle-hole channel is small, or (iii)
the total momentum $P=|{\bf p}+{\bf p}_1|$ is close to zero.
Outside these regions, contributions to the collision integral
appear to be minor.

In dealing with small momentum transfers, we first address the
scalar component $\Gamma_s$ of the scattering amplitude $\Gamma$,
which obeys the standard equation \cite{trio,lanl}
\beq
\Gamma_s(q,\omega)=f +f \Pi_0(q,\omega)\Gamma_s(q,\omega)\equiv
[f^{-1}-\Pi_0(q,\omega)]^{-1} \ ,
\label{grho}
\eeq
where $f$ is
the scalar part of the Landau interaction function. In FL theory,
the polarization loop $\Pi_0$ is an integral over the product of
two quasiparticle Green functions
$G(p,\varepsilon)=(\varepsilon-\xi(p))^{-1}$, given by
\beq
\Pi_0(q,\omega)=2 \int {n({\bf p})-n({\bf p}-{\bf q})\over
\xi({\bf p})-\xi({\bf p}-{\bf q})-\omega}d\upsilon \ ,
\label{pio}
\eeq
in which $d\upsilon=d^2p/(2\pi)^2$ is the volume element in
2D momentum space, and $n(p)=1/[1+\exp(\xi(p)/T)]^{-1}$ is the
quasiparticle momentum distribution.

The value of ${\rm Re}\,\Pi_0$ is of order of $N(0)$, the density
of states, proportional to $M^*$. In a strongly correlated  FL
obeying Landau theory, this quantity, whose sign depends on the
ratio $\omega/q$, is enhanced by the factor $M^*/M$ compared to
the corresponding ideal Fermi-gas value. On the other hand at
small $\omega$ and $q>q_{\mbox{\scriptsize min}}=M^*\omega/p_F$,
the imaginary part
of $\Pi_0(q,\omega,T=0)$, given by \cite{stern1}
\beq
{\rm Im}\,\Pi_0(q,\omega, T=0)\simeq
-\frac{\omega(M^*)^2}{\pi qp_F \sqrt{1-(M^*\omega/qp_F)^2}} \ ,
\label{impi2}
\eeq
has the same
order as ${\rm Re}\,\Pi_0$. Thus in strongly correlated systems
 $f^{-1}$ can be neglected, and  Eq.~(\ref{grho}) reduces to
 $|\Gamma_s(q\sim
q_c,\omega\sim T)|\simeq N^{-1}(0)$.\cite{schuck}

A similar situation applies for the spin-dependent part $\Gamma_a$
of the scattering amplitude $\Gamma$, which satisfies the same
equation (\ref{grho}) with the replacement $f\to f_a$, where $f_a$
is the spin-dependent part of the Landau interaction function. The
2D Fermi systems in question do not exhibit ferromagnetism, in
spite of the negative sign of $f_a$ derived from experimental data
on the spin susceptibility.  This means that the Pomeranchuck
stability condition \cite{pin,lanl} $1+f_aN(0)>0$ is not violated,
implying that $|f_aN(0)|<1$ holds even if the enhancement of the
effective mass is large.  The estimate $|\Gamma_a(q\sim
q_c,\omega\sim T)|\leq N^{-1}(0)$ follows straightforwardly.  In
the transverse particle-hole channel, where small momentum
transfer corresponds to $q\simeq 2p_F$, the situation  is
evidently the same, so that in the collision term (\ref{imsig}),
integration over $q$ can be restricted to the region of small $q$,
and the result is doubled.

In the third relevant momentum region  where the total momentum
$P$ is small, the scattering amplitude $|\Gamma(P\to 0)|\simeq -
1/[N(0)\ln P^2]$ contains an additional suppression factor $1/\ln
P^2$ due to the BCS logarithmic divergence of the
particle-particle propagator.\cite{lanl} Therefore in what
follows the respective contribution will be neglected. Thus we
conclude that a proper treatment of damping effects in the
strongly correlated system leads to substantial suppression of the
interaction factor $W$ governing the damping rate (\ref{imsig}).

In the foregoing analysis, the momentum dependence of the Landau
interaction function ${\cal F}=f+f_a{\bm \sigma}_1{\bm \sigma_2}$
has been neglected.  Upon its inclusion, relation (\ref{grho}) is
replaced by the integral equation
$$
\Gamma({\bf n} {\bf n}_1, q,\omega)={\cal F}({\bf n}{\bf n}_1)
$$
\beq
{}+\Pi_0(q,\omega)\!\int\! {\cal F}({\bf n}{\bf n}')\,
\Gamma({\bf n}' {\bf n}_1, q,\omega)\,d\varphi'/2\pi \ ,
\label{ints}
\eeq
with ${\bf n}={\bf p}/p_F$ and ${\bf n}_1={\bf p}_1/p_F$. We now
make use of the smallness of the quantity $1/\Pi_0\sim M/M^*$ and
rewrite the scattering amplitude as
\beq
\Gamma({\bf n} {\bf n}_1,q,\omega) =
X({\bf n}{\bf n}_1)/ \Pi_0(q,\omega) \ .
\label{gamx}
\eeq
Neglecting small corrections, Eq.~(\ref{ints})
becomes
\beq
0={\cal F}({\bf n}{\bf n}_1)+\int\! {\cal F}({\bf
n}{\bf n}')\,
 X({\bf n}'{\bf n}_1)\, d\varphi'/2\pi \ .
\label{intx}
\eeq
Inserting the expression (\ref{gamx}) into the collision
formula (\ref{imsig}), standard algebra \cite{quinn} converts
it to
$$ \gamma(\varepsilon\sim T) \sim \int
\limits_0^{\,\,\varepsilon\sim T}\!
\int\limits_{q_{\mbox{\scriptsize min}}}^{q_c}\!\!
\int X^2({\bf n}{\bf n}_1)\, {{\rm Im}\Pi_0(q,\omega)\over
|\Pi_0(q,\omega)|^2}
$$
\beq {}\times
 {\rm Im}\, G_R({\bf p} {-}{\bf q},\varepsilon{-}\omega)\,
d\omega\, q dq\, d\varphi\ ,
\label{imsig2}
\eeq
where ${\rm Im}G_R(p,\varepsilon)$ is given by Eq.~(\ref{imgr}).
In the 2D dilute
electron gas where $X=1$, this result practically coincides with
that  derived in Ref.~\onlinecite{quinn}. As usual,\cite{migdal}
integration over angles  in Eq.~(\ref{imsig2}) is replaced by
integration over energies
 $\xi(l)$ where ${\bf l}={\bf p}-{\bf q}$, and after some
algebra we arrive at \cite{quinn}
\beq
\gamma(T)\sim {T^2M^*\over M\varepsilon^0_F}\,
\ln\left({M^*T\over M\varepsilon^0_F}\right) \ .
\label{quin}
\eeq
The same estimate is valid for other strongly correlated 2D
Fermi systems where the spectrum $\xi(p)$ is specified only by
the effective mass.

The result (\ref{imsig2}) holds in the
density region where the effective mass $M^*$
diverges, since in its derivation only relation (\ref{gamx}) has
been  employed. Here at relevant $q,\omega$, the value of ${\rm
Re}\Pi_0(q,\omega)$ turns out to be of order
$(dp/d\xi)_{\xi\simeq T}$.
As for ${\rm Im}\,\Pi_0(q,\omega)$, its value is evaluated on
the base of the general
formula \cite{trio}
$$
{\rm Im}\,\Pi_0(q,\omega)=\!\int\!\!\int \left[
\tanh{\varepsilon-\omega\over 2T}-\tanh {\varepsilon\over 2T}\right]
$$
\beq
{}\times {\rm Im}\,G_R({\bf p}{-}{\bf q},\varepsilon{-}\omega)\,
{\rm Im}\,G_R({\bf p},\varepsilon)\,{d\varepsilon\over \pi}\,
d\upsilon \ .
\eeq
Insertion of the explicit form of ${\rm Im}\,G_R$ and integration
over $\xi(p),\xi(l)$ along the same lines, as before, gives
\beq
{\rm Im}\,\Pi_0(q,\omega)\sim{p_F\over q}\int
\left[
\tanh{\varepsilon{-}\omega\over 2T}{-}\tanh{\varepsilon\over 2T}
\right] \left({dp\over d\xi}\right)_T^2 d\varepsilon ,
\label{impo1}
\eeq
where the product $(dp/d\xi)_{\xi=\varepsilon}
(dp/d\xi)_{\xi=\varepsilon-\omega}$
has been replaced by $\left(dp/d\xi\right)_T^2$. As a result, one finds
\beq
|{\rm Im}\Pi_0(q,\omega{\sim}T)|\sim {Tp_F\over q}\!
\left({dp\over d\xi}\right)_T^2\ .
\label{impior}
\eeq
Upon inserting this result into Eq.~(\ref{imsig2}) we are led to
\beq
\gamma(T)\sim {T^2\over p_F} \left({dp\over d\xi}\right)_T
\ln\left[{T\over p_F}\left({dp\over d\xi}\right)_T\right] \ .
\label{gamsc}
\eeq

\begin{figure}[t]
\begin{center}
\includegraphics[width=0.7\linewidth,height=0.6\linewidth]{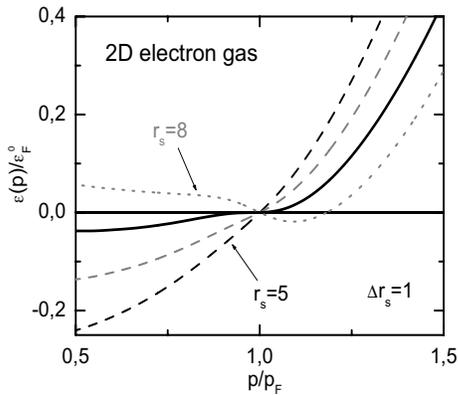}
\end{center}
\caption
{Single-particle spectrum $\xi(p)$ of the homogeneous
two-dimensional electron gas in units of $\varepsilon_F^0=p^2_F/2M$,
evaluated at $T=0$ for different values of $r_s=(\pi n)^{-1/2}/a_B$,
where $a_B$ is Bohr radius.
}
\label{fig:2deg}
\end{figure}

Thus for evaluation of the damping rate $\gamma(T)$ in the density region
where $M^*$ diverges one needs to
know the spectrum $\xi(p)$ close to the Fermi surface.
To date, microscopic calculations in this
density region have been performed only for the electron gas in 2D
and 3D and only at $T=0$.\cite{kss,zks,shag,dassarma1,dassarma2}
In Fig.~\ref{fig:2deg}, we display results for the spectrum $\xi(p)$
of the 2D electron gas, calculated at $T=0$ within a functional
approach \cite{physrep}
that successfully reproduces the ground-state energies of several
model Fermi systems that have been benchmarked by Monte Carlo
simulations. Close to the Fermi surface, the electron spectrum
$\xi(p,n_{\infty})$, given in Fig.~\ref{fig:2deg}, behaves as
$(p-p_F)^3$. The leading FL term re-emerges at finite
temperatures,\cite{shaghf} so that
\beq
\xi(p,T,n_{\infty})=p_F(p-p_F)/M^*(T,n_{\infty})+\xi_3(p-p_F)^3\ ,
\label{spct}
\eeq
with the effective mass \cite{kz1,prb} going
like $M^*(T,n_{\infty})\sim\left(dp/d\xi\right)_T\sim T^{-2/3}$.
With this result, the damping rate evaluated with the help of
Eq.~(\ref{gamsc}), becomes $\gamma(T,n_{\infty}) \sim T^{4/3}\ln
(\varepsilon^0_F/T)$.

\begin{figure}[t]
\begin{center}
\includegraphics[width=0.7\linewidth,height=1.0\linewidth]{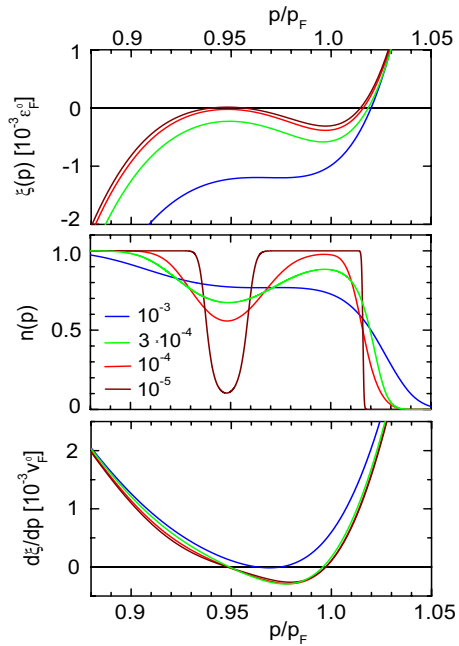}
\end{center}
\caption
{Single-particle spectrum $\xi(p)$ in units of
$10^{-3}\,\varepsilon_F^0$ (top panel), occupation numbers $n(p)$
(middle panel), and $d\xi/dp$ in units of $10^{-3}\,v_F^0$, where
$v_F^0=p_F/M$ (bottom panel), plotted versus $p/p_F$ at four
color-coded temperatures relevant to the bubble phase, in units of
$\varepsilon_F^0$. The model (\ref{mod0}) is assumed with
parameters $\beta_1=0.07$ and $\lambda_1=0.45\,N_0$, where
$N_0=p_FM/\pi^2$.}
\label{fig:0pfb}
\end{figure}
\begin{figure}[t]
\begin{center}
\includegraphics[width=0.7\linewidth,height=1.0\linewidth]{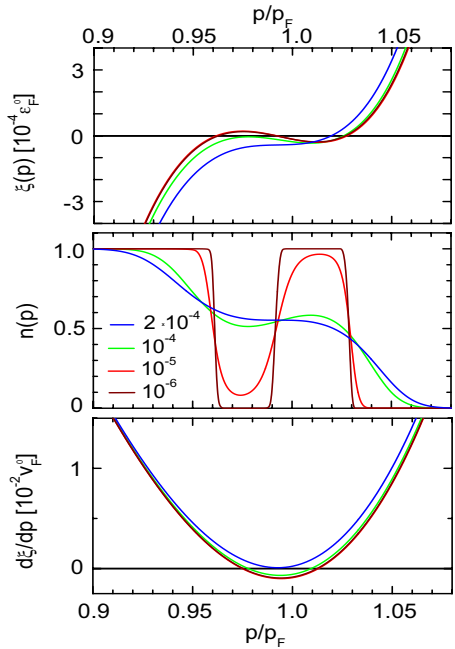}
\end{center}
\caption
{Same as in Fig.~\ref{fig:0pfb} but
for the model (\ref{model}) with parameters $\lambda_2=3\,N_0$ and
$\beta_2=0.48$. } \label{fig:2pfb}
\end{figure}

The critical single-particle spectrum $\xi(p{,}T{=}0{,}n_{\infty})\sim
(p-p_F)^3$ is not universal. In a broader context, the Landau
state is known (e.g.\ from Refs.~\onlinecite{zb,baldo}) to lose its
stability at a density $n_b$ for which a bifurcation point $p=p_b$
emerges in equation \beq \xi(p,T=0,n_b)=0\   , \label{root} \eeq
which ordinarily has only the single root $p=p_F$. The particular
form $\xi(p,T=0,n_{\infty})\sim (p-p_F)^3$ corresponds to the case
in which the bifurcation point $p_b$ coincides with $p_F$.
Obviously, in the general case one has $p_b\neq p_F$, and the
Landau state loses its stability before $M^*$ becomes infinite. If
the distance between $p_b$ and $p_F$ is small, then the
single-particle spectrum has the form \beq
\xi(p,T=0,n_b)\sim(p-p_b)^2(p-p_F) \ . \label{spnb} \eeq

Suppose the temperature $T$ lies below the maximum value $\xi_m$
of $|\xi(p,T=0,n_b)|$ in the momentum interval $[p_b,p_F]$.  In
this case, the dominating contributions to the properties of
interest come from the momentum region adjacent to the bifurcation
point $p_b$, where according to Eq.~(\ref{spnb}), $(dp/d\xi)_T\sim
T^{-1/2}$.  From this result and Eq.~(\ref{gamsc}) one obtains
$\gamma(T,n_b)\sim T^{3/2}\ln(\varepsilon^0_F/T)$.

 Beyond the critical density $n_b$,
Eq.~(\ref{root}) possesses two additional roots $p_1<p_b<p_2$.
The single-particle spectrum $\xi_{FL}(p,T=0,n)$, evaluated with
the Landau momentum distribution $n_{FL}(p)=\theta(p_F-p)$, has
the form \beq \xi_{FL}(p,T=0,n)\sim (p-p_1)(p-p_2)(p-p_F) \ .
\label{root1} \eeq If $p_b\neq p_F$, the roots $p_1,p_2$ are both
located either in the interior of the Fermi sphere or both outside
it.  If $p_b=p_F$, then $p_1<p_F<p_2$.  In all these cases, the
Landau occupation numbers $n_L(p)$ are rearranged. As a rule, the
Fermi surface becomes multi-connected, but the quasiparticle
occupation numbers $n(p)$ continue to take values 0 or 1.  Hence
the Landau-Migdal quasiparticle picture holds, with $n(\xi)=1$ for
$\xi<0$ and 0 otherwise.  Consider first the case $p_1<p_2<p_F$.
Then according to Eq.~(\ref{root1}), the single-particle states
remain filled in the intervals $p<p_1$ and $p_2<p<p_F$, while the
states corresponding to $p_1<p<p_2$ are empty. We call this new
phase the bubble phase. If the bifurcation point $p_b$ coincides
with the Fermi momentum $p_F$, then $p_1{<}p_F$ and $p_2{>}p_F$,
and the states with $p{<}p_1$ and with $p_F{<}p{<}p_2$ are
occupied, while those for $p_1{<}p{<}p_F$ are empty.  Again one
deals with the bubble phase.

\begin{figure}[t]
\begin{center}
\includegraphics[width=0.95\linewidth,height=0.9\linewidth]{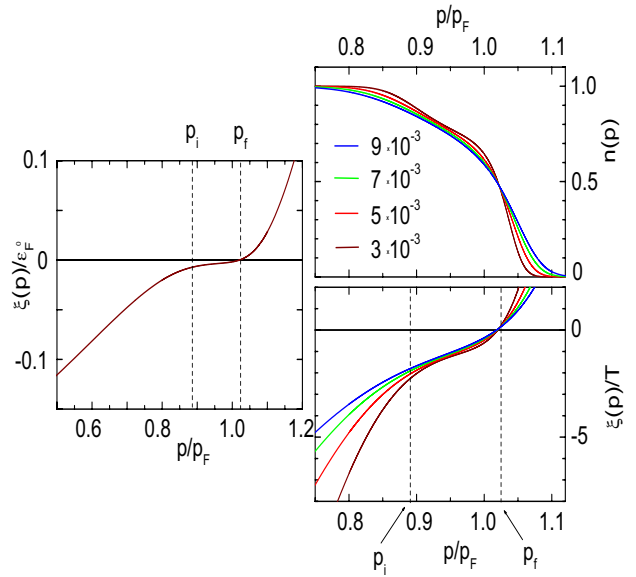}
\end{center}
\caption
{Single-particle spectrum $\xi(p)$ in units of
$\varepsilon_F^0$ at the critical temperature $T_Z=3\cdot
10^{-3}\,\varepsilon^0_F$ (left panel), occupation numbers $n(p)$
(right-top panel), and $\xi(p)/T$ (right-bottom panel), plotted
versus $p/p_F$ at four color-coded temperatures relevant to the
phase with a fermion condensate, in units of $\varepsilon_F^0$.
The model (\ref{mod0}) is assumed.} \label{fig:0pffc}
\end{figure}

\begin{figure}[t]
\begin{center}
\includegraphics[width=0.95\linewidth,height=0.9\linewidth]{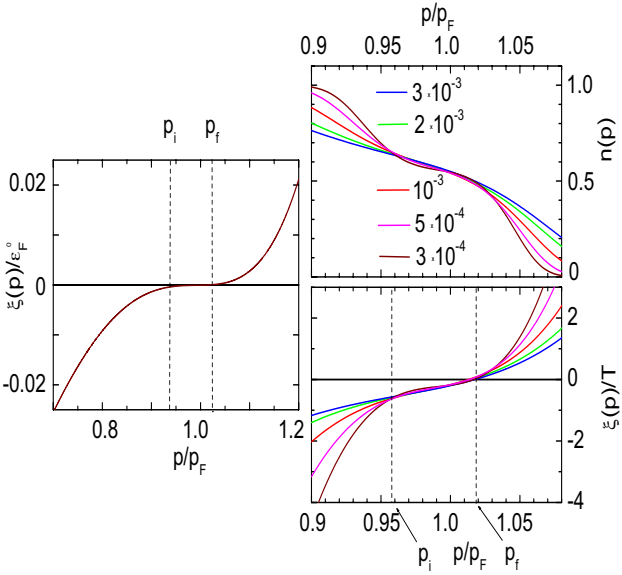}
\end{center}
\caption
{Same as in Fig.~\ref{fig:0pffc} but for the model
(\ref{model}). The single-particle spectrum in the left panel is
shown at $T_Z=3\cdot 10^{-4}\,\varepsilon^0_F$. }
\label{fig:2pffc}
\end{figure}
At this point, we observe that the solution (\ref{root1}) is not
self-consistent, since the spectrum is evaluated with $n_{FL}(p)$
while the true Fermi surface is doubly-connected. Following
Ref.~\onlinecite{zb}, we consider the feedback of the rearrangement of
$n_{FL}(p)$ on the spectrum $\xi(p)$ in the bubble phase based on
the Landau relation \cite{trio,lanl}
\beq
{\partial\epsilon(p)\over\partial {\bf p}}={{\bf p}\over M}
+ \int f({\bf p},{\bf p}_1) {\partial n(p_1)\over
\partial {\bf p}_1}d\upsilon_1 \ ,
\label{lansp}
\eeq
where, as before, $f$ is
the scalar part of the Landau interaction function and
$n(p)=[1+\exp(\xi(p)/T)]^{-1}$ is the quasiparticle momentum
distribution.  Solutions of this nonlinear integral equation are
known only in 3D Fermi systems with phenomenological  functions
$f$ depending on  $q=|{\bf p}-{\bf p}_1|$.  Despite of the
diversity of forms assumed for  $f(q)$, the resulting
single-particle spectra and momentum distributions bear a close
family resemblance.   This robustness is illustrated in
Figs.~\ref{fig:0pfb} and \ref{fig:0pffc}, which display results
\cite{zb} from solution of Eq.~(\ref{lansp}) for the  function
\beq
f(q)=\lambda_1/[(q/p_F)^2+\beta_1^2] \ ,
\label{mod0}
\eeq
and in Figs.~\ref{fig:2pfb} and \ref{fig:2pffc}, which present
results for \beq f(q)=\lambda_2/[((q/2p_F)^2-1)^2+\beta_2^2]\ .
\label{model} \eeq

Let us briefly summarize how solutions of Eq.~(\ref{lansp}) evolve
under variation of $T$. When the bubble range $p_2-p_1$ is small,
then heating to $T\sim T_{FL}=(p_2-p_1)^2/M$ results in its
dissolution (see Figs.~\ref{fig:0pfb} and \ref{fig:2pfb}).
 With further increase of $T$, the function $\xi(p)$ becomes
smoother, and in the region of a new critical temperature $T_Z$, a
flat portion $\xi\simeq 0$ appears in the spectrum over an
interval $[p_i,p_f]$ surrounding the Fermi momentum $p_F$, as
shown in the left panels of Figs.~\ref{fig:0pffc} and
\ref{fig:2pffc}. Since $\xi(p)=\epsilon(p)-\mu$ and
$\epsilon(p)=\delta E_0/\delta n(p)$, the equality $\xi=0$ can be
rewritten as a variational condition \cite{ks}
\beq
{\delta E_0\over \delta n(p)}=\mu \ , \quad p_i<p<p_f \ ,
\label{var}
\eeq
with $E_0=\sum_{{\bf p}}\epsilon^0_{{\bf p}}n({\bf p})+{1\over 2}
\sum_{{\bf p},{\bf p}_1}f({\bf p}-{\bf p}_1)\, n({\bf p})\, n({\bf p}_1)$
and $\epsilon^0_{\bf p} = p^2/2M$.  The solution $n_0(p)$
of Eq.~(\ref{lansp}), or equivalently of Eq.~(\ref{var}), is a
continuous function of $p$ with a nonzero derivative $dn_0/dp$
(see Figs.~\ref{fig:0pffc} and \ref{fig:2pffc}, top-right panels).
The set of
single-particle states with $\xi(p)=0$ is called the fermion
condensate (FC), since the corresponding density of states
$\rho(\varepsilon)$ contains a Bose-liquid-like term $\eta
n\delta(\varepsilon)$. The dimensionless constant $\eta\simeq
(p_f-p_i)/p_F$ is naturally identified as a characteristic
parameter of the FC phase.

It has been demonstrated \cite{noz} that the FC ``plateau'' in
$\xi(p)$ has a small slope, evaluated by inserting $n_0(p)$ into
the above Fermi-Dirac formula for $n(\xi)$ to yield \beq
\xi(p,T\geq T_Z)=T\ln {1-n_0(p)\over n_0(p)} \,, \,\, p_i<p<p_f \,
. \label{spt} \eeq As indicated in the bottom-right panels of
Figs.~\ref{fig:0pffc} and \ref{fig:2pffc}, at  $T\geq T_Z$ the
ratio $\xi(p)/T$ is indeed a $T$-independent function of $p$ in
the FC region. The presence of this flat portion of $\xi(p)\sim T$
is a signature of the phenomenon called fermion
condensation.\cite{ks,noz,vol} The width
$\xi(p_f)-\xi(p_i)\equiv \xi_f-\xi_i$
of the FC ``band'' appears to be of order $T$, almost
independently of $\eta> \eta_{\mbox{\scriptsize min}}\sim
10^{-2}$. Thus at $\eta>\eta_{\mbox{\scriptsize min}}$ the FC
group velocity is estimated as  \beq \left({d\xi(p)\over
dp}\right)_T\sim {T \over \eta p_F} \ , \quad p_i<p<p_f \ .
\label{estfc} \eeq

As shown in Ref.~\onlinecite{shag}, the effective mass diverges
 before attaining the critical point
$r_{\mbox{\scriptsize CDW}}$ for the charge-density-wave
instability.
Microscopic calculations confirm this assertion: $M^*$ diverges at
$r_s=r_{\infty}\simeq 7$, while the condensate of the
charge-density waves occurs at $r_{\mbox{\scriptsize CDW}}\simeq 10$.
Thus in the interval
$r_{\infty}<r_s<r_{\mbox{\scriptsize CDW}}$ one deals with
the homogeneous ground state having a FC.
In this case, the damping rate of single particle excitations
is evaluated  on the base of the formulas (\ref{gamsc}) and
(\ref{estfc}) that yields
\beq
\gamma(T)\sim \eta\, T \ln (1/\eta) \ .
\label{bcrfc}
\eeq
Thus,
as long as the FC density remains small, the dimensionless damping
rate $r(T)=\gamma(T)/T$ proves to be small as well, so the
presence of the FC does not destroy the quasiparticle picture. It
is worth noting that at greater energies $\varepsilon\gg T$, the
damping $\gamma(\varepsilon)$ grows as $\sqrt{\varepsilon}$ with
increasing $\varepsilon$.\cite{khz} Thus at these energies, the
ratio $\gamma(\varepsilon)/\varepsilon$ exceeds unity, and the
quasiparticle picture fails independently of the $\eta$ value.

Let us now apply our results to the elucidation of behavior of the
conductivity of the dilute 2D electron gas  from data obtained in
samples with silicon inversion layers.\cite{krav,shash} At
densities $n>2 \cdot 10^{11} {\rm cm}^{-2}$, this behavior is
explained within traditional screening
theory,\cite{stern2,dolg,dassarma3,dassarma4} taking
into account only
impurity contributions. However, at lower densities, the 2D
electron gas undergoes a metal-insulator transition (MIT), as
signaled by a change in sign of the derivative $d\rho(T\to 0)/dT$.
In high-quality samples, the sign change occurs at the density
$n_{\mitr}\simeq 0.9\cdot 10^{11} {\rm cm}^{-2}$. On the metallic
side of the MIT, this derivative has positive sign, while on the
insulating side, it is negative, the separatrix
$\rho_{\mitr}(T)\simeq 3h/e^2\simeq 75$\,k$\Omega$ between the two
phases being almost horizontal.\cite{krav,shash}

 At these densities,
the electron-electron interaction, first taken into account within
perturbation theory in Ref.~\onlinecite{zala}, becomes a ``play maker''.
A crucial point is that close to the critical density $n_{\mitr}$,
the effective mass $M^*(n)$ diverges.\cite{krav,shash} In this
situation, a standard method of treatment of kinetic phenomena on
the  base of the Boltzmann equation fails. Therefore we employ a
different approach where
 the conductivity $\sigma(T)$ is expressed in terms of the
imaginary part of the polarization operator $\Pi({\bf j},\omega\to
0,T)$ through \cite{trio}
\beq
\sigma(T)\sim -\lim_{\omega\to 0}\omega^{-1}\,
{\rm Im }\,\Pi({\bf j},\omega,T) \ .
\label{cond1}
\eeq
It provides contributions of two different types, namely
from (i) imaginary parts of the quasiparticle Green functions and
(ii) imaginary parts of the scattering amplitudes. As a rule, both
these contributions provide the same $T$-dependence of
$\sigma(T)$. E.g. this is seen  from homogeneous systems without
impurities where the two types of contributions cancel each other
to ensure vanishing of the resistivity due to momentum
conservation. (In solids, the resistivity $\rho(T)$ differs from 0
due to umclapp processes). Such cancellation allows one  to find
out  the $T$-dependence of the resistivity in the  critical
density  region, where the spectrum $\xi(p)$  becomes flat by
retaining in Eq.~(\ref{cond1}) only contributions coming from ${\rm
Im}\,G_R$. Thereby the calculations are simplified considerably,
and the expression for ${\rm Im}\,\Pi$ acquires the form
$$
 {\rm Im}\,\Pi({\bf j},\omega{\to}0,T)\sim \!\!\int\!\!\!\int
\left(
\tanh{\varepsilon{-}\omega\over 2T}{-}\tanh {\varepsilon\over 2T}
\right)
$$
\beq
{}\times |{\cal T}({\bf j},\omega{=}0)|^2\,
{\rm Im}\,G_R({\bf p},\varepsilon{-}\omega)\,
{\rm Im}\,G_R({\bf p},\varepsilon)\, d\varepsilon\, d\upsilon \ ,
\eeq
\label{imp2}
Here, ${\cal T}$ is the vertex part, whose static limit is given
by \cite{trio,lanl}
${\cal T}({\bf j},\omega=0)=e\partial\xi(p)/\partial {\bf p}$.
Upon inserting the
explicit form for ${\rm Im}\,G_R$ into Eqs.~(\ref{imp2}) and
(\ref{cond1}), the latter becomes \cite{vorugani}
\beq
\sigma(T)=2e^2\!\!\int\!\!\!\int
{(d\xi/dp)^2\,\gamma^2(\varepsilon)\,d\varepsilon\, d\upsilon
\over 2T [(\varepsilon{-}\xi(p))^2{+}\gamma^2(\varepsilon)]^2
\cosh^2({\varepsilon/2T})} \ .
\label{imp3}
\eeq
Converting, as before, the momentum integration to an integration
over $\xi$ and
taking into account that the overwhelming contributions to this
integral come from the vicinity of the point $\xi=\varepsilon$, we
arrive at \beq \sigma(T)=2 \pi {ne^2\over p_F} \int{(d\xi/dp)
\over 2T\,\gamma(\xi)\cosh^2({\xi/2T})}\, d\xi \ , \label{cond3}
\eeq where $n=p^2_F/2\pi$. Remembering that in the region of the
critical density  of 2D electron gas, where the effective mass
diverges, one has  $\gamma(T)\sim T^{4/3}\ln(\varepsilon_F^0/T)$,
and then Eq.~(\ref{cond3}) gives us
\beq
\sigma(T)\sim T^{-2/3}/\ln(\varepsilon_F^0/T) \ .
\label{met}
\eeq
Since $d\sigma(T\to 0)/dT<0$, this point is
situated on the metallic side of the MIT. Beyond this density,
i.e. at $r_{\infty}<r_s<r_{\mbox{\scriptsize CDW}}$, and greater
but still very low temperatures $T\sim T_Z$, we pass the point of
fermion condensation. The FC contribution to $\sigma(T)$ is
evaluated with the help of Eqs.~(\ref{bcrfc}) and (\ref{estfc}),
yielding
\beq
\sigma_{FC}(T)=\sigma_0 e^2/\eta^2\ln (1/\eta) \ ,
\label{condf}
\eeq
where $\sigma_0$ is a $T$-independent constant.

At $r_s>r_{\mbox{\scriptsize CDW}}$,  the spontaneous generation
of the condensate of
the charge density waves occurs, and the ground state of 2D
electron gas becomes nonhomogeneous. Consequently, a gap  in the
single-particle spectrum opens that results in exponential falling
of the conductivity $\sigma(T)$ at $T\to 0$, implying that one
deals with  the insulating side of the MIT. Thus the separatrix,
dividing the metallic and insulating domains, is situated in the
FC region, and according to Eq.~(\ref{condf}), it  is a straight
line. This result is in agreement with available experimental
data.\cite{krav,vitk,shash}

\begin{figure}[t]
\begin{center}
\includegraphics[width=0.7\linewidth,height=0.5\linewidth]{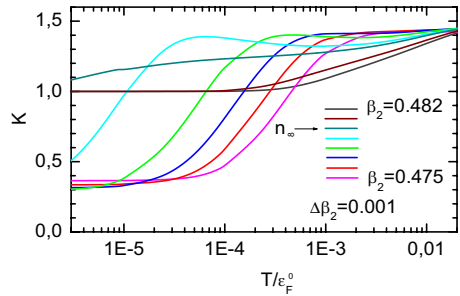}
\end{center}
\caption
{The dimensionless Hall coefficient $K=R_H\,nce$, plotted versus
$T/\varepsilon_F^0$, evaluated in the model (\ref{model}) at $H=0$
with $\lambda_2=3\,N_0$ and different color-coded values of
the parameter $\beta_2$.
}
\label{fig:hall2pf}
\end{figure}

Flattening of the single-particle spectrum entails the change of
the Hall coefficient $R_H=\sigma_{xyz}/\sigma^2_{xx}$.\cite{norman}
In  homogeneous matter at $H\to 0$, $\sigma_{xx}=\sigma/3$,
with $\sigma$  given by Eq.~(\ref{cond3}),
while $\sigma_{xyz}$ is recast to
\beq
\sigma_{xyz}={e^3\over 3c\gamma^2}\int \left({d\xi\over dp}\right)^2
{\partial n(\xi)\over \partial \xi} d\xi \ ,
\label{sxyz}
\eeq
where $n(\xi)$ is the Fermi-Dirac distribution function.
Far from the critical density $n_{\infty}$,  these formulas lead
to the standard result $R_H=1/nec$. The critical spectrum
 of 2D electron gas has the form $\xi(p,n_{\infty},T=0)\sim
(p-p_F)^3$, and with the help of Eqs.~(\ref{cond3}), (\ref{sxyz}),
one then finds $R_H=K/nec$ where
\beq
K(n_{\infty},T\to 0, H\to 0)=
{\int z^{4/3}e^z[1+e^z]^{-2}dz\over
\left(\int z^{2/3}e^z[1+e^z]^{-2}dz\right)^2}\simeq 1.5 \ .
  \label{crit}
  \eeq
We see that at the critical density,
the effective volume of the Fermi sphere considerably shrinks.
It is important that  even quite close to the critical point where
the effective mass  still remains finite, the value  $K=1$
holds, so that at low $T$, the critical behavior (\ref{crit}) of
$K$ emerges abruptly. In Fig.~\ref{fig:hall2pf}, we display results
for the coefficient $K$ as a function of temperature, calculated
at $H=0$ in the model (\ref{model}) at different values
of the parameter $\beta_2$. On the other hand, imposition of
static magnetic field $H$ on the system at the critical density
$n_{\infty}$ renders the effective mass finite \cite{shaghf,prb}
and hence, one can expect the abrupt change of the Hall
coefficient $R_H(n_{\infty},T\to 0,H)$ as a function of $H$.

In conclusion, we have analyzed damping effects in the strongly
correlated 2D Fermi liquid in a density region where the effective
mass diverges.  We have demonstrated that in spite of the
enhancement of the dimensionless constants specifying the strength
of the effective interaction between quasiparticles, the
Landau-Migdal quasiparticle picture is applicable on both sides of
the phase transition associated with the divergence of the
effective mass.  The results of the analysis have been applied to
the interaction-driven metal-insulator transition in the 2D
electron gas, demonstrating that the separatrix between the
metallic and insulating regions is a straight line.

We thank V.~T.~Dolgopolov, V.~M.~Galitski, A.~A.~Shashkin,
and V.~M.~Yakovenko for valuable discussions. This research was
supported by NSF Grant PHY-0140316 (JWC and VAK), by the McDonnell
Center for the Space Sciences (VAK), and by the Grant
NS-1885.2003.2 from the Russian Ministry of Education and Science
(VAK and MVZ).

\end{document}